\documentclass{aa}
\usepackage{graphicx}
\begin{document}
\title{~~~~~~Photoionization models of roundish galactic \protect{\newline}planetary nebulae in the thick disk\thanks{Based
on observations collected at the European Southern Observatory, La
Silla under prog-id. 56.D-0717 and on observations made with the
European Southern Observatory telescopes obtained from the
ESO/ST-ECF Science Archive Facility.}}

\subtitle{}

\author{
M. Emprechtinger\inst{1}\and T. Rauch\inst{2,3}\and S.
Kimeswenger\inst{1} }

\offprints{M. Emprechtinger,\\
\email{martin.emprechtinger@uibk.ac.at}}

\institute{Institut f{\"u}r Astrophysik der Universit{\"a}t
Innsbruck, Technikerst. 25, A-6020 Innsbruck, Austria \and
Dr.-Remeis-Sternwarte, Sternwartstra{\ss}e 7, D-96049 Bamberg,
Germany \and Institut f\"ur Astronomie und Astrophysik, Abteilung
Astronomie, Sand 1, D-72076 T\"ubingen, Germany }

\date{Received / accepted}

\abstract{We present the result of photo-ionizing modelling of the
three planetary nebulae (PNe) A~20, A~15 and MeWe~1-3. All three
objects are roughly roundish, highly excited and have a high
galactic $z$. The PNe displayed low densities in the shell, but
relatively dense halos. A low metallicity and a relative high
electron temperature were found. Comparisons with radio
observations confirmed the obtained properties. The objects are
very likely originating from thick disk stellar progenitors. The
distances found investigating the PNe shells are somewhat lower
than those derived spectroscopically with the central stars in the
past.

}
\titlerunning{Photoionization models of roundish galactic planetary nebulae in the thick disk}
\maketitle


\section{Introduction}

Although the interacting wind model for the formation of planetary
nebulae (PNe) is accepted since a long time (Kwok et
al.~\cite{kwok}), the exact mechanism is not clearly understood up
to now. Several numerical simulations have been carried out
especially during the last decade (e.g. Frank \&
Mellema~(\cite{frank2}), Dwarkadas \& Balick~(\cite{dwar}) or
Sch\"onberner \& Steffen~(\cite{schon}). The last one carried out
an one-dimensional hydrodynamic simulation based on the
evolutionary tracks of Bl\"ocker~(\cite{blocker}) and the
simulations of the circumstellar shell around Asymptotic Giant
Branch (AGB) star (Steffen
et al.~\cite{steffen}). \\
The results of their simulation showed that a shock-bounded
ionized main PN shell moving supersonically into the ABG-material
compressing the inner parts of the matter into a dense but thin
shell. The unaffected AGB-material becomes ionized as well and
forms a rapidly expanding halo. Due to the drop of luminosity of
the central stars of the PNe (CSPN) towards the white dwarf regime
the outer part of the PN shell recombines quickly forming an
second inner halo. The disadvantage of this simulation is that it
just considers one dimension. Thus more complex interactions of
the winds like Rayleigh-Taylor instabilities, quickly building a
clumpy environment as we see it in HST images, are not taken into
account. This causes that the filling factor ($\rm\varepsilon$)
was set in fact to unity. Comparing densities obtain from
forbidden line ratios and the one obtained from radio data or
similar methods clearly show a filling factor far below unity
(e.g. Mallik \& Peimbert~\cite{fill2} and Boffi \&
Stanghellini~\cite{fill3}). This changes the dramatically the
optical thickness for ionizing UV photons and thus a recombining
PN-shell is doubtful (Armsdorfer et al. \cite{birgit}).

\noindent The three PNe A~20, A~15  (Abell~\cite{Abell}) and
\mbox{MeWe~1-3} (Melmer \& Weinberger~\cite{MeWe}) were selected
as they are roughly roundish objects. The central stars of the PNe
(CSPNe) are well studied spectroscopically (Rauch et
al.~\cite{rauch1}, McCarthy et al.~\cite{McC}, Saurer et
al.~\cite{saurer}). This allows us to fix the input of radiation
for the nebula. We modelled the physical properties of PNe-shells
of three objects and compare the results with the theory.

\begin{figure*}[ht]
\centerline{\resizebox{18cm}{!}{\includegraphics{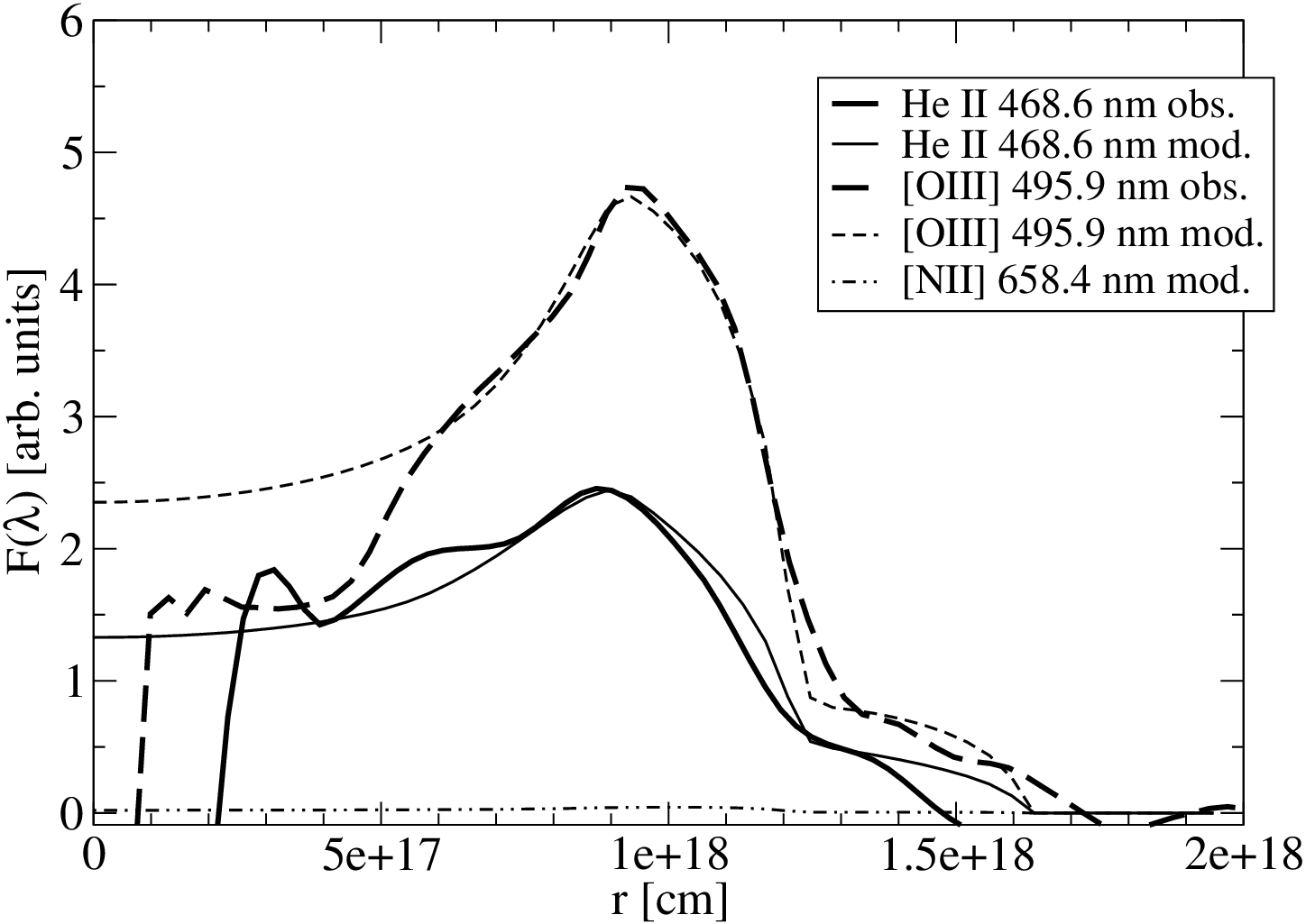}\phantom{lkfag}\includegraphics{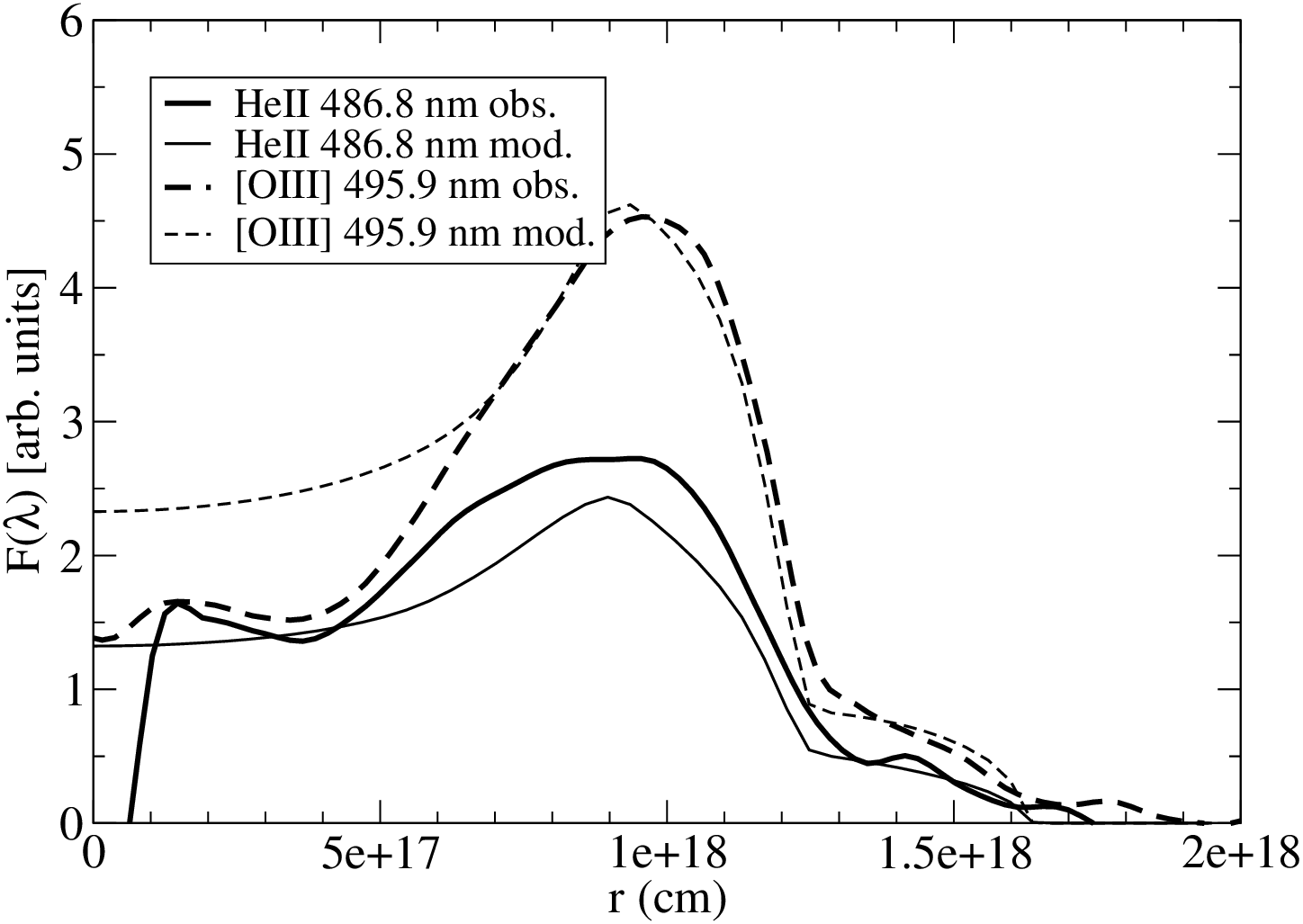}}}

\caption{The observations and the model for several lines of A 20.
\noindent{\bf{Left:}}\ Observation with NTT/EMMI (slit
N--S);~\noindent{\bf{Right:}}\ Observation with
\mbox{ESO~3.6~m/EFOSC} (slit E--W); especially in the [OIII] lines
the radial symmetry of the object is well pronounced. Also the
fragmentation in three parts, halo, PN shell and central cavity,
is displayed well. The [NII] lines are in an upper limit and thus
confirm clearly the low metallicity found here (see Tab.
\ref{chem214}).} \label{mod214a}
\end{figure*}

\section{Input Data}
\subsection{Observations and  Data Reduction}

Most data were obtained 1992 January 29$^{\rm th}$ and January
30$^{\rm th}$ using  the New Technology Telescope (NTT) with the
EMMI spectrograph at ESO La Silla, Chile and were taken from the
ESO data archive. THX31156 (red arm) or TK1024AF (blue arm) CCDs
and grating \#8 were used. The slit-width was adjusted to
1\farcs2. The spatial resolution was $0\farcs44$ per pixel at the
red arm of the instrument (480 nm to 680 nm) and $0\farcs37$ at
the blue arm (400 nm to 505 nm). Additionally A~20 was observed by
one of us (TR) in 1996 February 10$^{\rm th}$ with the ESO 3.60\,m
telescope using the EFOSC1 spectrograph and the \#26 TEK512M
1215-3 detector at a resolution of $0\farcs61$ per pixel (from 380
nm to 530 nm). At EFOSC observations narrow band
[\ion{O}{III}] and \ion{H}{$\alpha$} images was taken too.\\
The calibration was done using usual procedure in MIDAS and the
calibration data from EMMI manual and from EFOSC1 manual. The
standard star EG 21 (Hamuy et al. \cite{hamuy}) was
used for flux calibration.\\
The target selection for this  investigation followed two
criteria: First, the PN should have a roughly round shape; second,
the CSPN should be spectroscopically studied.

\begin{table}[ht]
\caption{Basic data for the PNe sample:} \label{obj}
\begin{tabular}{lccc}
\hline\noalign{\smallskip}
Name & GPN  & F$_{\rm 1.4~GHz}$ & IRAS \\
 & {\scriptsize (Kimeswenger \cite{KS})} & [mJy] &  \\
\hline \hline\noalign{\smallskip}
{\tt A 20} & {\tt G214.96$+$07.81} & 10.4 & {\tt 07203$+$0151} \\
{\tt A 15} & {\tt G233.53$-$16.31} & $<2.5$ & {\tt 06249$-$2520} \\
{\tt MeWe 1-3} & {\tt G308.26$+$07.79} & - & {\tt 13249$-$5426} \\
\hline
\end{tabular}
\end{table}

For all three objects IRAS counterparts have been found and
radio-observations at 1.4 GHz have been found for two of them
(Condon \& Kaplan~\cite{con}). The objects are summarized in
Tab.~\ref{obj}.

\begin{figure*}[ht]
\centerline{\resizebox{18cm}{!}{\includegraphics{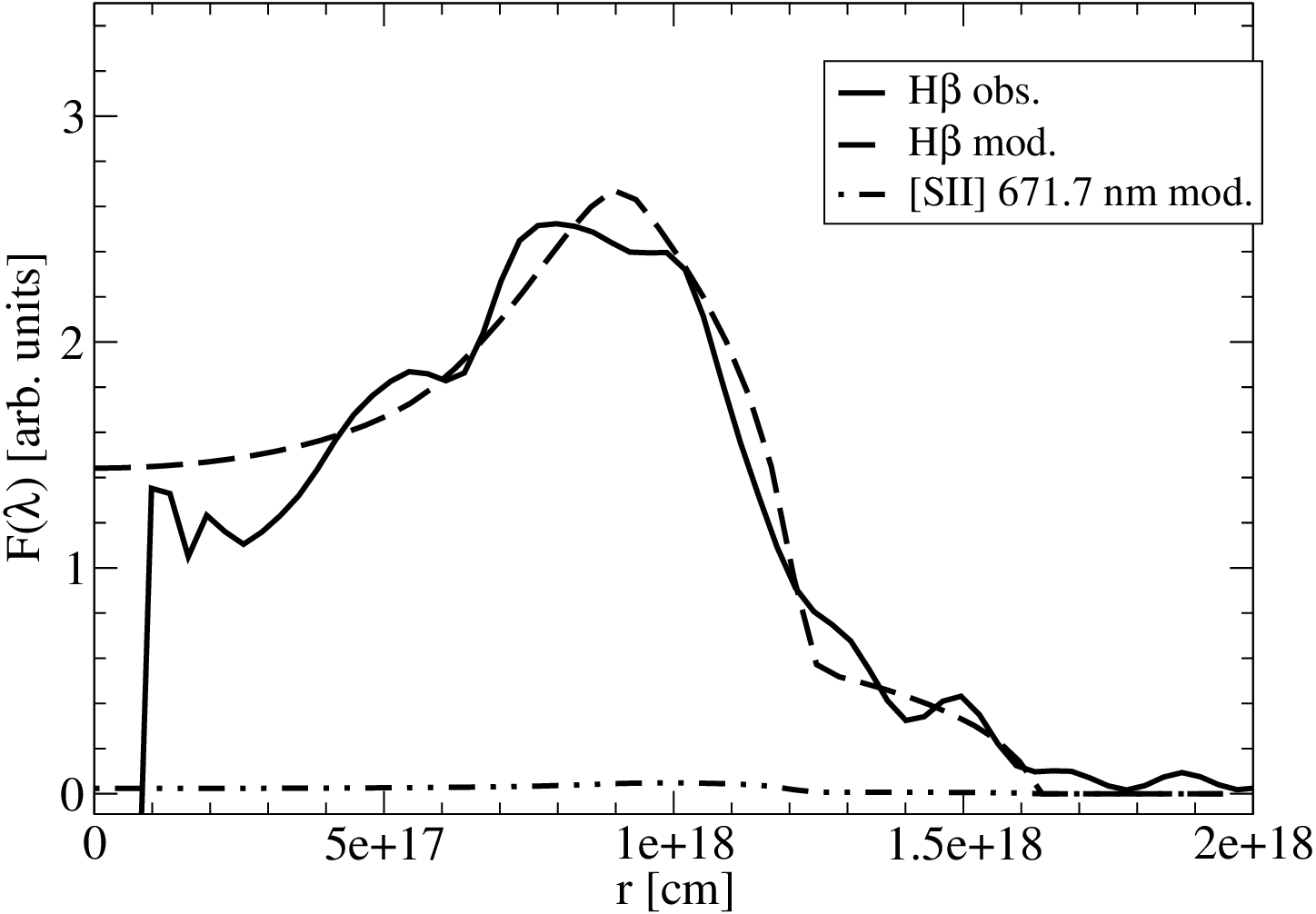}\phantom{lkfag}\includegraphics{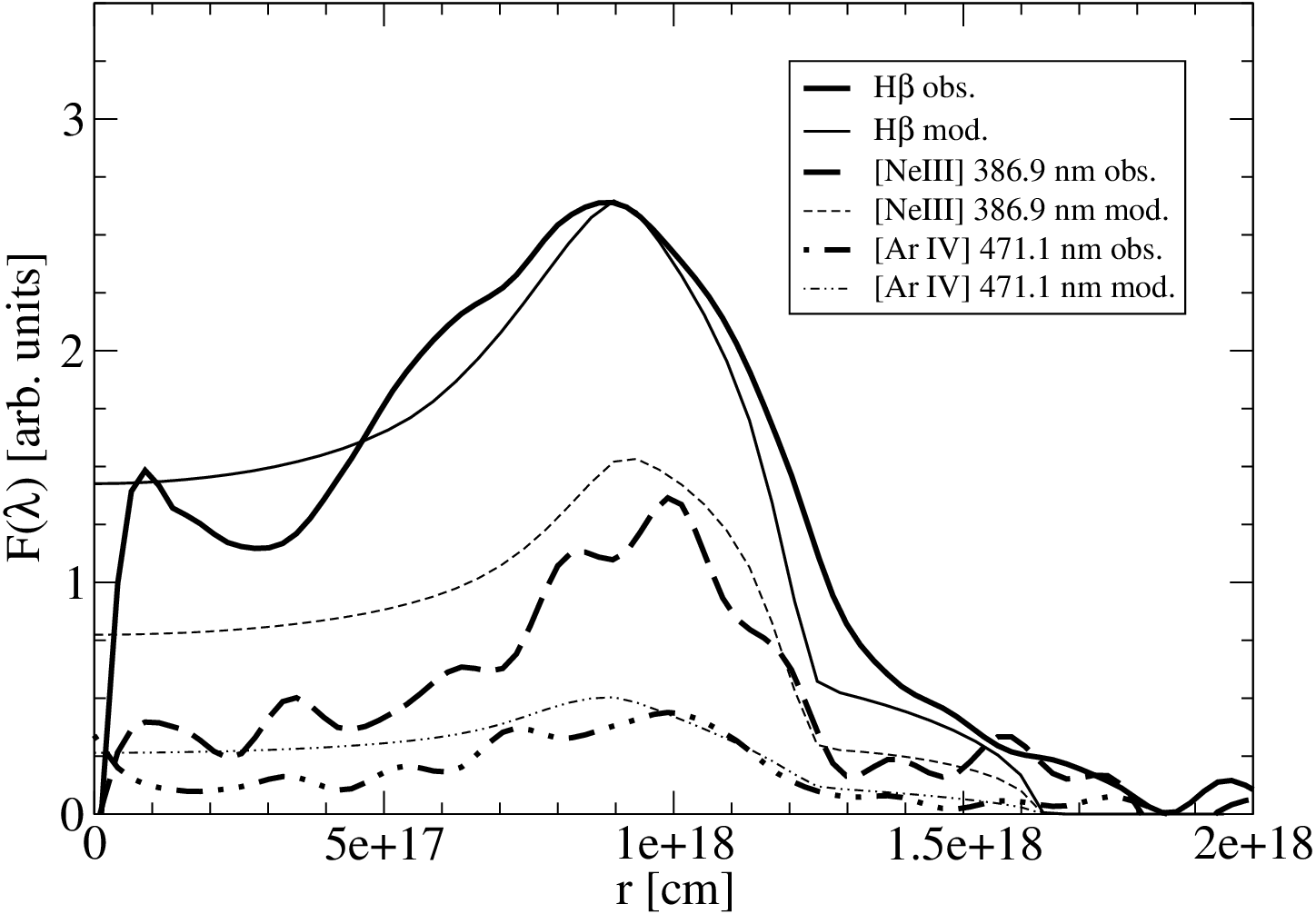}}}

\caption{Comparison of the observation and the model of several
lines of A 20 (same as in Fig.~\ref{mod214a}). } \label{mod214b}
\end{figure*}

\subsection{Spectroscopy}

All three objects are characterized by an intensive HeII line
($\rm\lambda=468.3~nm$) and a total depletion of [NII] and [SII]
lines. The [OIII] lines at $\rm\lambda=495.9~nm$ and
$\rm\lambda=500.7~nm$ are much weaker than expected as well. The
extinction was derived from the Balmer-decrement and to deredden
the frames we used the interstellar extinction law by Savage \&
Mathis~(\cite{Sava}). The results are given in Tab.~\ref{extinct}.
For A15 and A20 our results from the Balmer-decrement of the high
S/N spectra differ significantly from those derived from the
survey spectra in Tylenda et al. (\cite{tyl92}). After dereddening
with our values from Tab.~\ref{extinct} the continuum of the CSPN
fits perfectly to our results. Also the $B-V$ of the CSPN from
literature supports our results. The resulting $(B-V)_0$ is then
always between -0\fm34 and -0\fm25. This is expected for the hot
CSPNe. The relative errors of the line ratios were estimated by
the variations of the ratios of $\rm H\alpha/H\beta$ and $\rm
[OIII]_{500.7}/[OIII]_{495.9}$ along the radius. A conservative
estimation gives an error of 10-15\% for the relative line
strength. Since the three [OIII] lines at $\rm\lambda=500.7~nm,
\lambda=495.9~nm$ \& $\rm\lambda=436.3~nm$ are detected a
temperature determination was possible (Tab.~\ref{temp}). The
error of the ratio of [OIII] lines was assumed to be 50 \%,
because the $\rm [OIII]_{436.3}$ was very faint. Hence the
accuracy of $\rm T_e$ was about $\rm T_{-3\,000}^{+11\,000}~K$.
Although the error of $\rm T_e$ is relative high it turned out,
that the electron temperature is significantly higher than the
$10\,000$ K, typical for PNe. This is a signature of the low
metallicity (see below), causing the cooling by forbidden lines to
fail.
\begin{table}
\caption{Interstellar foreground extinction from the Balmer
decrement derived here and from literature:} \label{extinct}
\begin{tabular}{lcc}
\hline\noalign{\smallskip}
Name & ~~~E(B-V)~~~ & E(B-V) \\
     & here & ~~~Tylenda et al. (\cite{tyl92}) \\
\noalign{\smallskip}\hline \hline\noalign{\smallskip}
A 15 & 0\fm00 & 0\fm04 $-$ 0\fm64 \\
A 20 & 0\fm05 & 0\fm30 $-$ 0\fm35 \\
MeWe 1-3 & 0\fm27 & --- \\
\hline
\end{tabular}
\end{table}\begin{table}
\caption{Electron temperatures derived from [OIII] lines and
computed in CLOUDY:} \label{temp}
\begin{tabular}{lccc}
\hline\noalign{\smallskip}
Name & $\frac{I_{495.9}+I_{500.7}}{I_{436.3}}$ & $T_{e} [K]$ & $\left<{T_{e}^{\tt CLOUDY}}\right> [K]$\\
\noalign{\smallskip}\hline \hline\noalign{\smallskip}
\smallskip
A 15 & 54.7 & $16\,800_{-3\,000}^{+11\,000}$ & 23\,000 \\
\smallskip
A 20 & 71.5 & $14\,800_{-3\,000}^{+11\,000}$ & 20\,000 \\
\smallskip
MeWe 1-3 & 62.2 & $15\,800_{-3\,000}^{+11\,000}$ & 22\,500 \\
\hline
\end{tabular}
\end{table}

\subsection{Central stars}

The central stars of the objects have been studied very well in
the past.  Detailed spectroscopies were carried out by Rauch et
al.~(\cite{rauch1}) (A~20), McCarthy et al.~(\cite{McC}) (A~15)
and Saurer et al.~(\cite{saurer}) (MeWe~1-3). To determine
distances $D_{CSPN}$ the method of Heber et al.~(\cite{heber}) was
used for A~20 and A~15 in the papers above. In the case of
MeWe~1-3 no distance was published before. We thus did it using an
Eddington flux $H_\nu$ given by Rauch~(\cite{rauch4}) following
the method of Heber et al.~(\cite{heber}), who gave calibrations
for Johnson and Stroemgren filters.

\begin{table}[ht]
\caption{} \label{CSPN}
\begin{tabular}{lcccc}
\hline\noalign{\smallskip}
Name & $\rm T_{eff}$  & L & log g  & $D_{CSPN}$ \\
 &  [kK] & [$\rm L_{\odot}$]&  [$\rm cm/s^2$] &  [kpc] \\
\hline \hline\noalign{\smallskip}
A 20 & $119\pm 22$ & 2100 & 6.13 &$4.1\pm 50\%$  \\
A 15 & $110\pm 10$ & 4238 & 5.7 &$4.5\pm 25\%$  \\
MeWe 1-3 & $100\pm 10$ & 4580 & 5.5 &$6.9\pm 50\%$  \\
\hline
\end{tabular}
\end{table}

Unfortunately no reliable photometry is available for the CSPNe of
A~20 and MeWe~1-3, thus the error is considerably higher. But also
for A~15 an error of 25 \% was assumed according Mendez et
al.~(\cite{mendez}).

\section{Photoionisation Model}

To model the PNe the CLOUDY code (Ferland \cite{cloudy}) was used.
As input-parameters the radial density-profile, the chemical
abundances and the stellar properties (within the ranges of error
given above) were varied. The filling factor was assumed to be
0.1, which is typical for evolved PNe
(Mallik \& Peimbert~(\cite{fill2}) and Boffi \& Stanghellini~(\cite{fill3})).
The NLTE central star
models of Rauch (\cite{rauch2}, \cite{rauch3}) were used to obtain
a proper ionizing spectrum. Armsdorfer et al. (\cite{birgit})
showed the need of real NLTE stellar atmospheres to describe
especially the HeII lines properly.

This results in the fraction of ionization of hydrogen and helium
as a function of radius and a mean electron temperature. To obtain
the surface-brightness, integration over the emissivity along the
line of sight was carried out. To do this a spherical symmetry was
adopted, what is indicated by round shape of the objects (see
Soker~(\cite{soker1}) and Tylenda et al.~(\cite{Tyl}) for A~20 and
A~15). The surface-brightness of various emission lines have been
compared with the observation to achieve the best result. The
average electron temperature given by CLOUDY was compared with
those derived before by the [OIII] lines. Due to the weak 436.3~nm
line a spatial resolved study wasn't possible here.

 The objects A~15 and
MeWe~1-3 displayed enhancements in the southern regions,
especially in [OIII]. This might be due to interaction with the
interstellar medium. Thus this regions where not taken into
account for the model.

\subsection{A~20 (GPN G214.96$+$07.81)}

A~20 is the only one of these objects that is observed with both
telescopes, NTT and \mbox{ESO 3.60~m}. The advantage of the
doubled observation is that the slit direction was north-south at
NTT and east-west at the \mbox{ESO 3.60~m} telescope and thus we
were able to rule out systematic effects in the data reduction and
due to the assumed spherical symmetry.\\
In Fig.~\ref{mod214a} \&~\ref{mod214b} the result of the modelling
can be seen. The model was optimized with respect to the NTT
spectra. In the east-west direction the HeII line is slightly
stronger than in north-south but both observed spectra agree
fairly well with the model. Just in the very inner part the model
displays an emissivity, especially of the [OIII] line, that is too
high compared with the observations. This deficiency is, although
uncertain, most likely originating in deviations from the
spherical symmetry assumed in the CLOUDY code. The modelled [SII]
and [NII] lines are far below the limit of detection and thus
confirmed by the observation. EFOSC-images do not display the
regime, where the [NII] and [SII] lines are located, but the line
of [NeIII] at $\rm\lambda=386.9~nm$ and of [ArIV] at
$\rm\lambda=471.1~nm$. The [ArIV] line is very faint and the
[NeIII] at the blue limit of the spectrograph, where the
efficiency of the CCD and the transparency of the spectrograph
have not good performances. Thus their error is higher than the
one of the other lines.

\begin{figure}[ht]
\centerline{\resizebox{8.8cm}{!}{\includegraphics{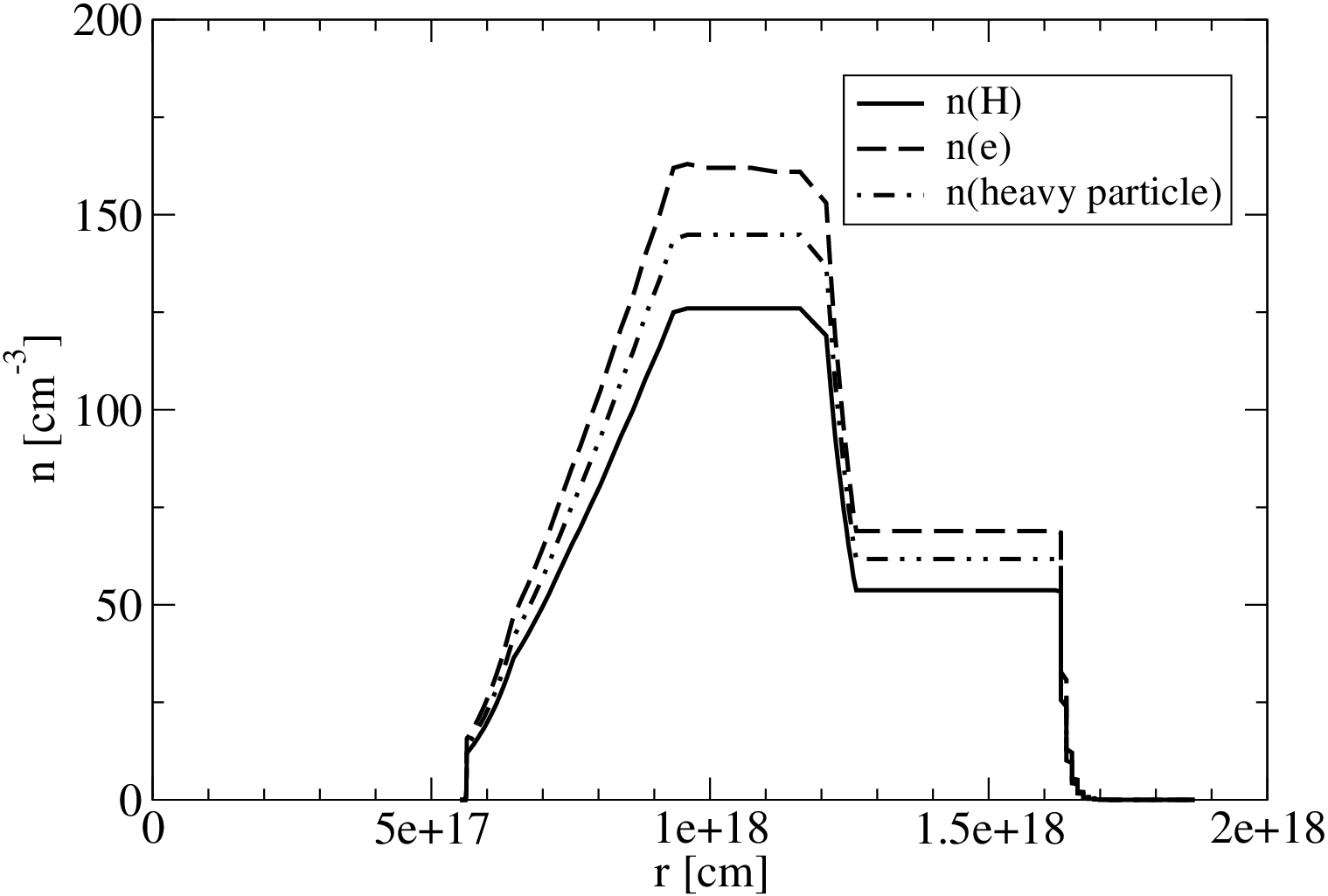}}}

\caption {Density profile of A~20. The hydrogen- and the electron
density are computed by CLOUDY. The heavy particle density is
calculated from the hydrogen density and the chemical
composition.}  \label{den214}
\end{figure}

\begin{figure}[ht]
\centerline{\resizebox{8.8cm}{!}{\includegraphics{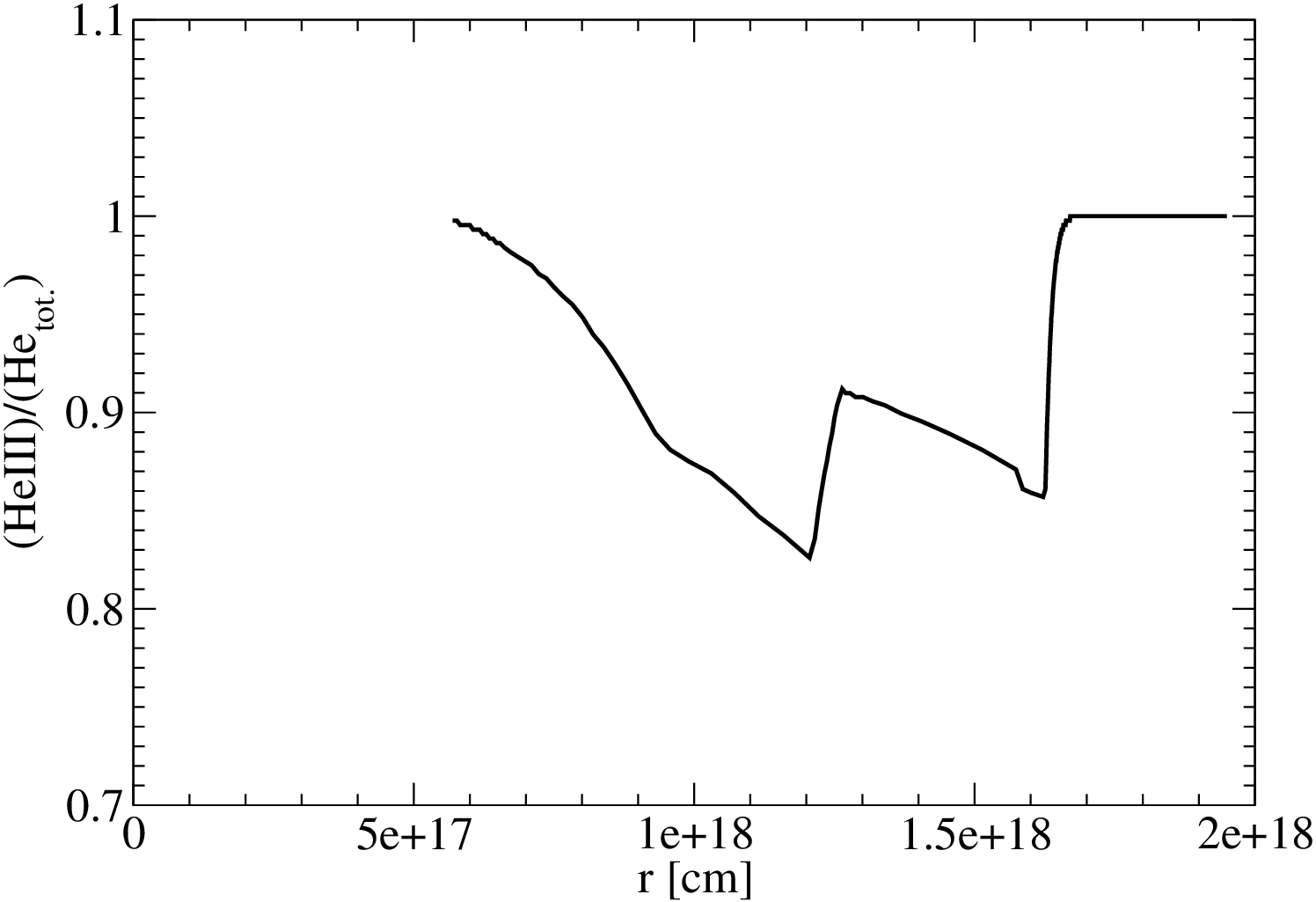}}}

\caption{The fraction of doubly ionized helium as a function of
radius.  As can be seen helium is mostly doubly ionized, as it was
expected considering the high $\rm HeII~468.6~nm$ intensity.}
\label{IO214}
\end{figure}

In Fig.~\ref{den214} the density profile of A~20 can be seen. The
heavy  particle density is mostly hydrogen and helium. Other
elements contribute less than $2\cdot 10^{-4}$. The high electron
density, which exceeds the hydrogen density twice as much as the
heavy particle density, indicates a mostly doubly ionized helium,
what is confirmed in Fig.~\ref{IO214}.

The chemical composition (Tab.~\ref{chem214}) shows clearly  an
underabundance of heavy elements by a factor of 10. This makes it
most likely to a member of the galactic thick disk population. The
total depletion of [SII] and [NII] lines can be explain just
partial by the underabundance of these elements. It is also caused
by the fact that these elements are mainly in a higher ionization
level. The overabundance of He is typical for PNe after the
dredge-up processes during the RGB/AGB phase.

The halo of the nebula is clearly highly ionized. It is not a
recombination halo as predicted by 1D hydro simulations.

\begin{table}[ht]
\caption{Chemical composition of A 20.  The abundances of each
element $Z$ are given in $\log\frac{[Z]}{[H]}$ and the
comparison with the solar abundances (Grevesse et al.
\cite{abund}) is given in $dex$.} \label{chem214}
\begin{tabular}{lccc}
\hline\noalign{\smallskip}
& abundance & abundance compared & dominating \\
& & to solar abundance & ionization level \\
\hline \hline\noalign{\smallskip}
He & -0.825 & 0.175 & He III \\
C & -4.45 & -1.00 & C IV \\
N & -5.00 & -0.97 & N IV \\
O & -4.10 & -0.97 & O IV \\
Ne & -4.824 & -0.90 & Ne IV \\
Mg & -5.523 & -1.10 & Mg III \\
Si & -5.523 & -1.07 & Si III \\
S & -5.824 & -1.15 & S IV \\
Ar & -5.80 & -0.32 & Ar IV \\
\hline
\end{tabular}
\end{table}

\subsection{A~15 (GPN G233.53$-$16.31)}

\begin{figure*}[ht]
\centerline{\resizebox{18cm}{!}{\includegraphics{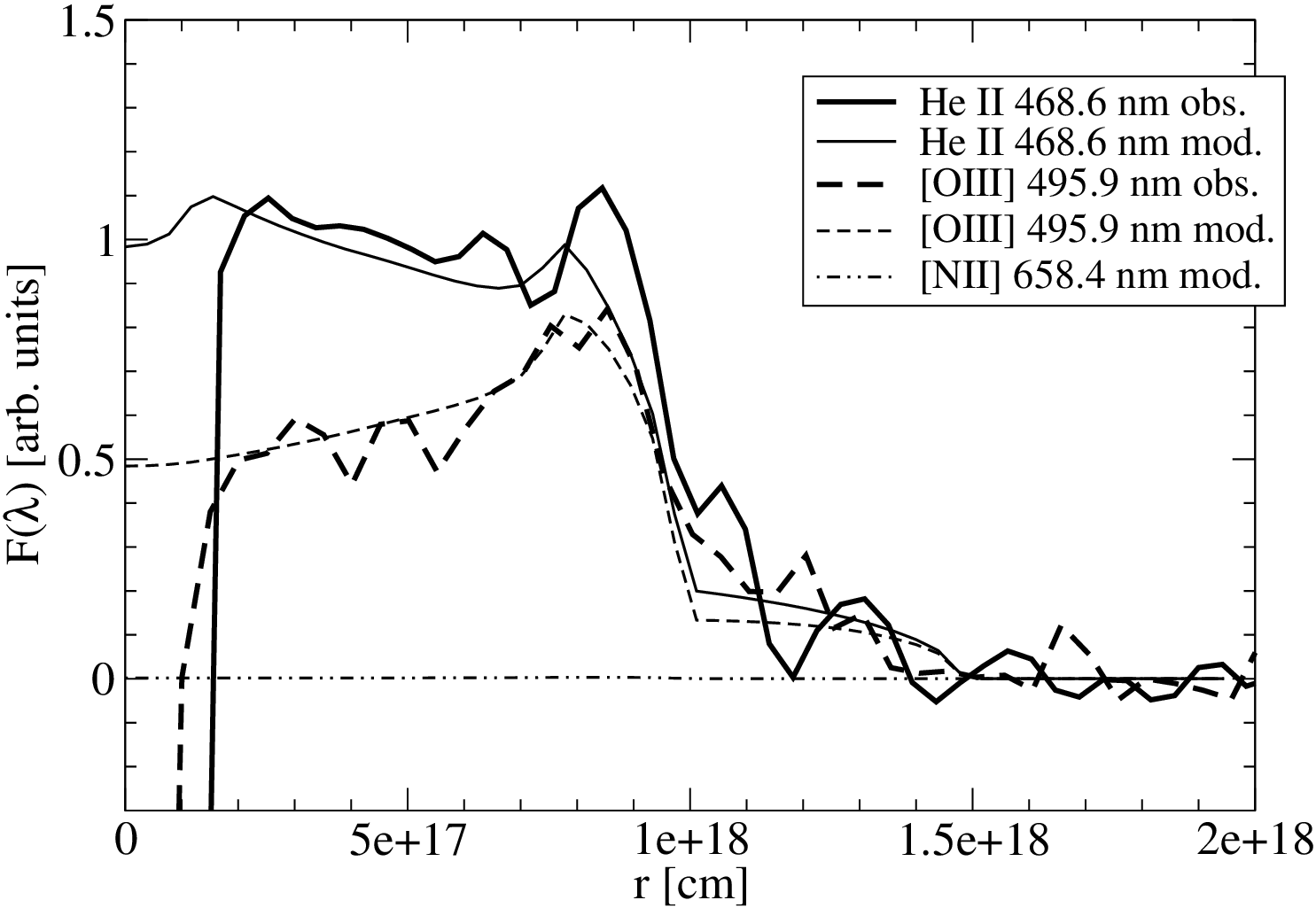}\phantom{lkfag}\includegraphics{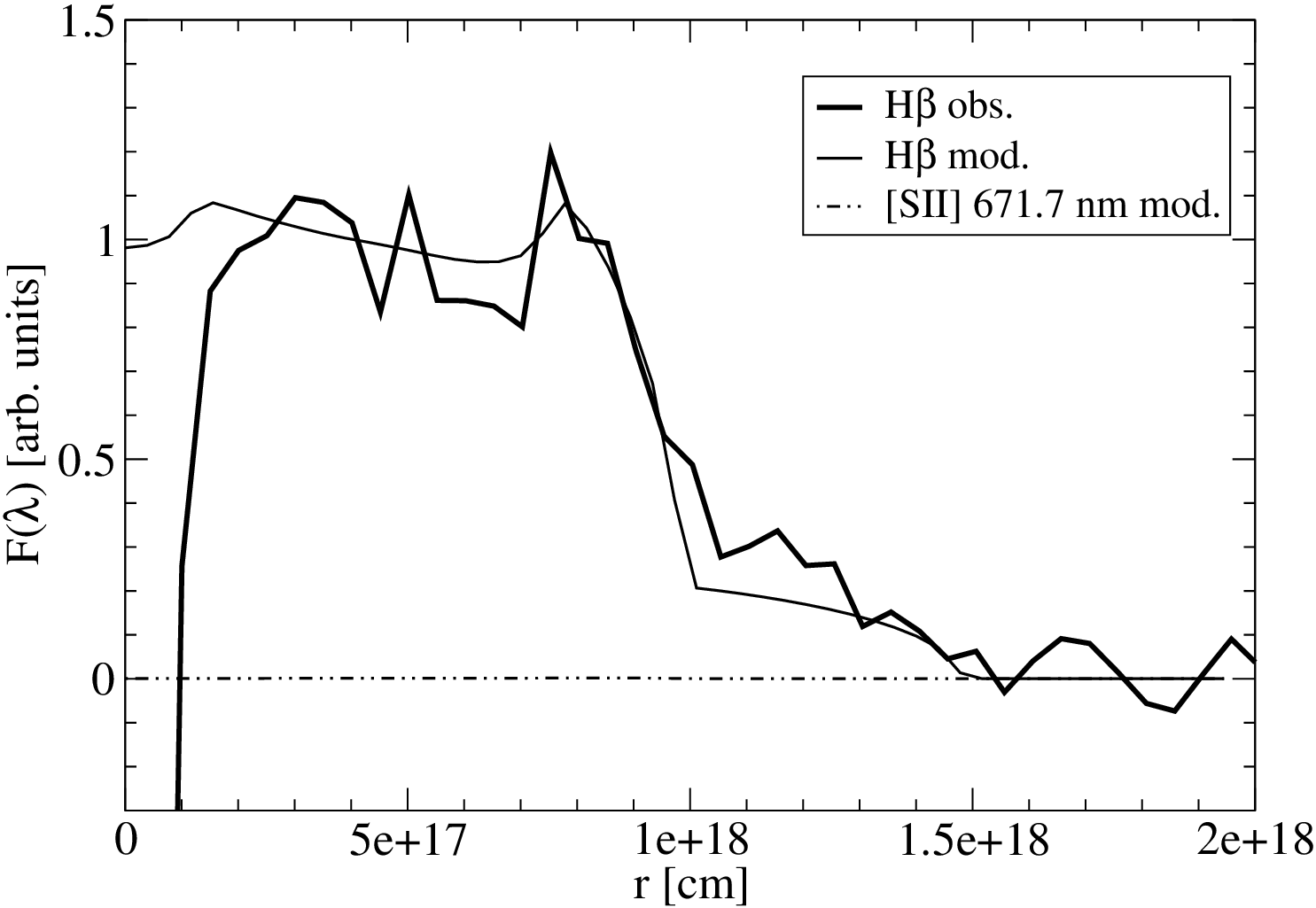}}}

\caption{Comparison of the observation and the model  of several
lines of A 15. $H\alpha$ and [OIII] ($\lambda=500.7\rm~nm$) are
not taken into account since they are related to H$\beta$ and
[OIII] ($\lambda=495.9\rm~nm$)  respectively. The [NII] and the
[SII] was not observable. The innermost 'hole' in the data is
faked due to the CSPN subtraction/overlay.} \label{moda15}
\end{figure*}

\begin{figure*}[ht]
\centerline{\resizebox{18cm}{!}{\includegraphics{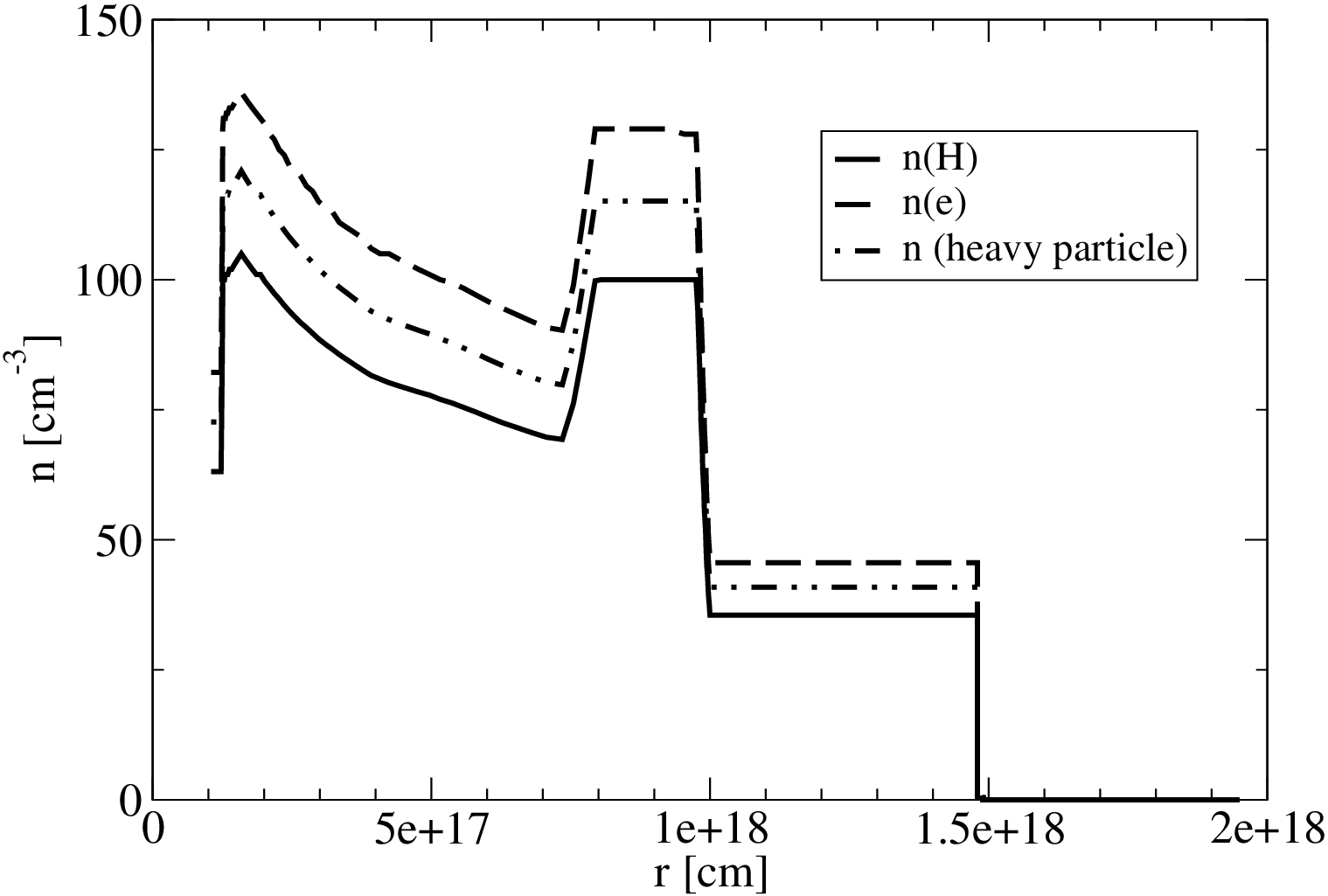}\phantom{lkfag}\includegraphics{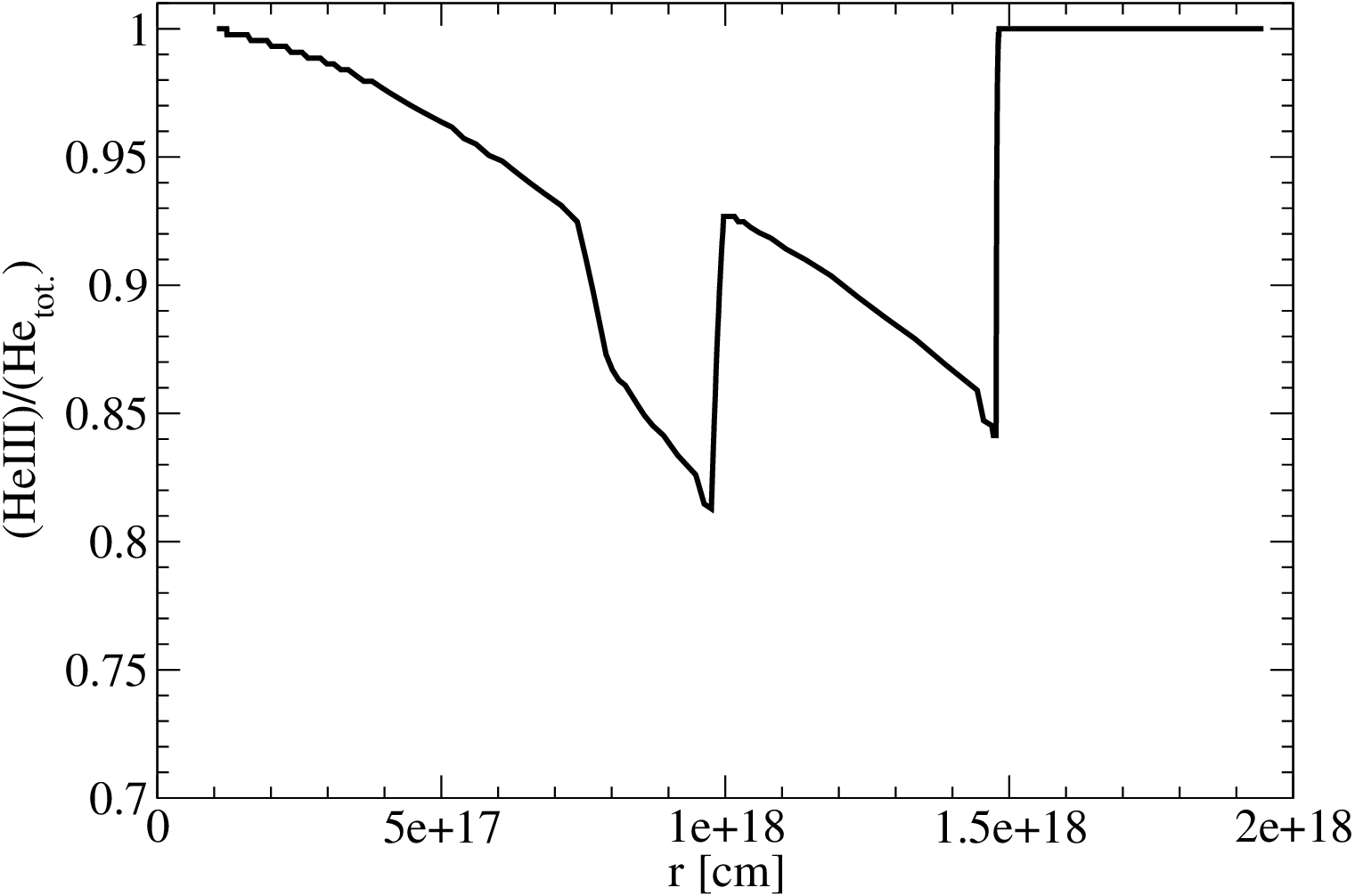}}}

\caption{Density profile and fraction of doubly ionized helium of
A15. Description like in Fig.~\ref{den214} \& Fig.~\ref{IO214}}
\label{dena15}
\end{figure*}

In Fig.~\ref{moda15} the surface brightness distributions for
several lines of  \mbox{A 15} are displayed. There is hardly any
decrease of the intensity towards the center of the nebula,
contrary to \mbox{A 20}. This indicates that the inner part of the
nebula is not empty, like in \mbox{A 20}, but somehow filled as
can be seen in Fig.~\ref{dena15}. The intensity of [SII] and [NII]
is again, in conformity with the observations, below the threshold
of detection. Also here the HeII line looks like the $\rm H\beta$.
Another remarkable fact is, that the [OIII] is less
prominent than the $\rm H\beta$ and the HeII line.\\
In the inner part of the nebula the density relates to the radius
as $r^{-0.3}$.The main shell is thinner and has a lower density
than the one of  \mbox{A~20} and followed by a relative extended
and dense halo.  The hydrogen in  \mbox{A 15} is also totaly
ionized as well as most of the helium (Fig.~\ref{dena15}).

The fraction of heavy elements is in this object  even slightly
lower than in \mbox{A 20} and the electron temperature thus is
higher as the cooling by the forbidden lines fails (see
Tab.~\ref{temp} what still agrees with the $T_e$=23000K computed
by CLOUDY). Also the dominating ionization levels are slightly
higher than the one in \mbox{A 20}, especially the one of C, S,
and Si. The typical He enrichment is shown by \mbox{A 15} too. The
overabundance of nitrogen with respect to the other heavy elements
give us not an hint for a massive progenitor, but originates in
the upper limit of the observed line.

\begin{table}[ht]
\caption{Chemical composition of A 15 (same as in
Tab.~\ref{chem214}).} \label{chema15}
\begin{tabular}{lccc}
\hline\noalign{\smallskip}
& abundance & abundance compared & dominating \\
& & to solar abundance & ionization level \\
\hline \hline\noalign{\smallskip}
He & -0.82 & 0.18 & He III \\
C & -4.95 & -1.50 & C V \\
N & -5.03 & -1.0 & N IV \\
O & -4.60 & -1.47 & O IV \\
Ne & -5.420 & -1.50 & Ne IV \\
Mg & -5.920 & -1.50 & Mg III \\
Si & -5.950 & -1.5 & Si V \\
S & -6.17 & -1.5 & S V \\
Ar & -6.98 & -1.5 & Ar IV \\
\hline
\end{tabular}
\end{table}

\subsection{MeWe 1-3 (GPN G308.26$+$07.79)}

\begin{figure*}[ht]
\centerline{\resizebox{18cm}{!}{\includegraphics{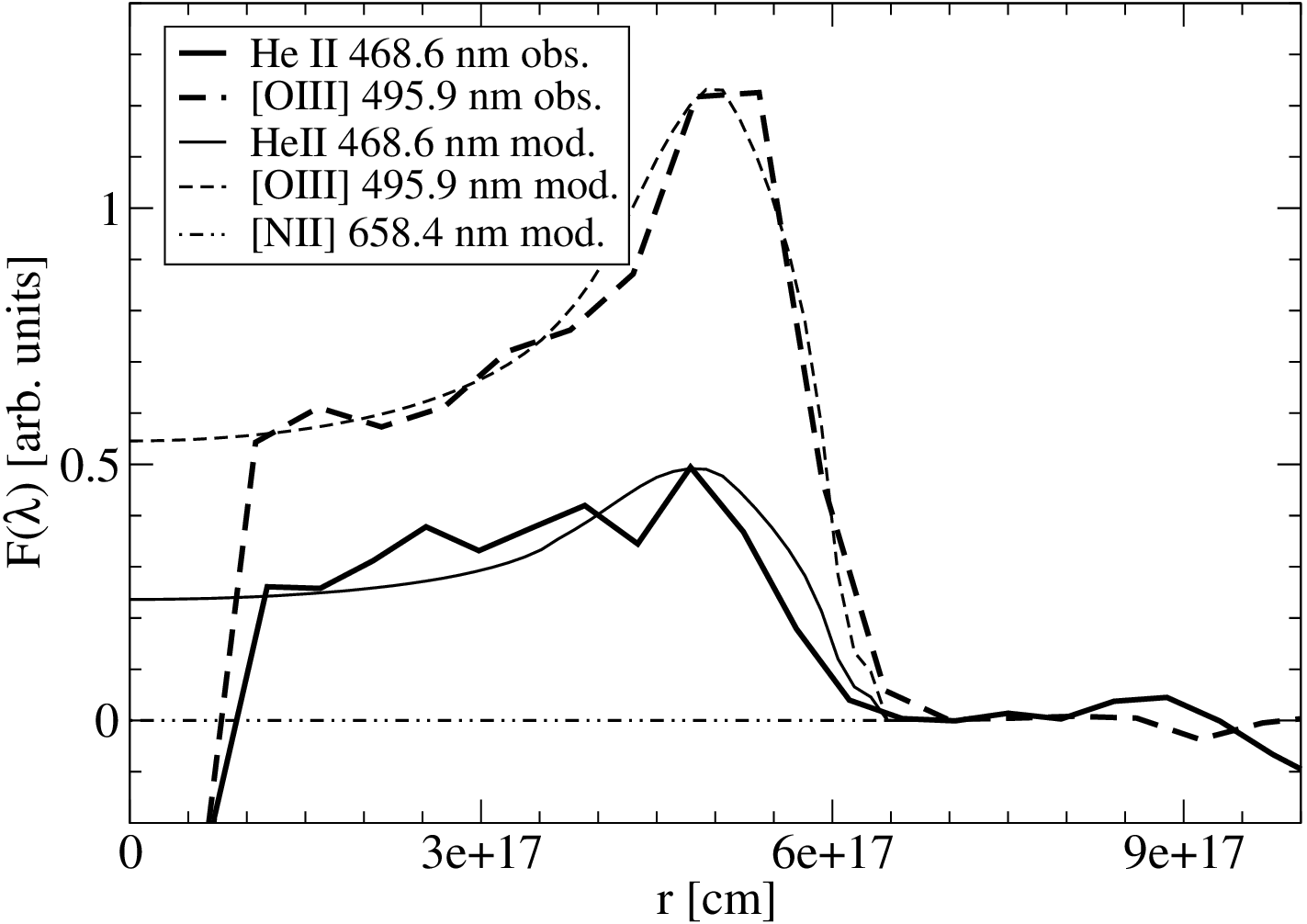}\phantom{lkfag}\includegraphics{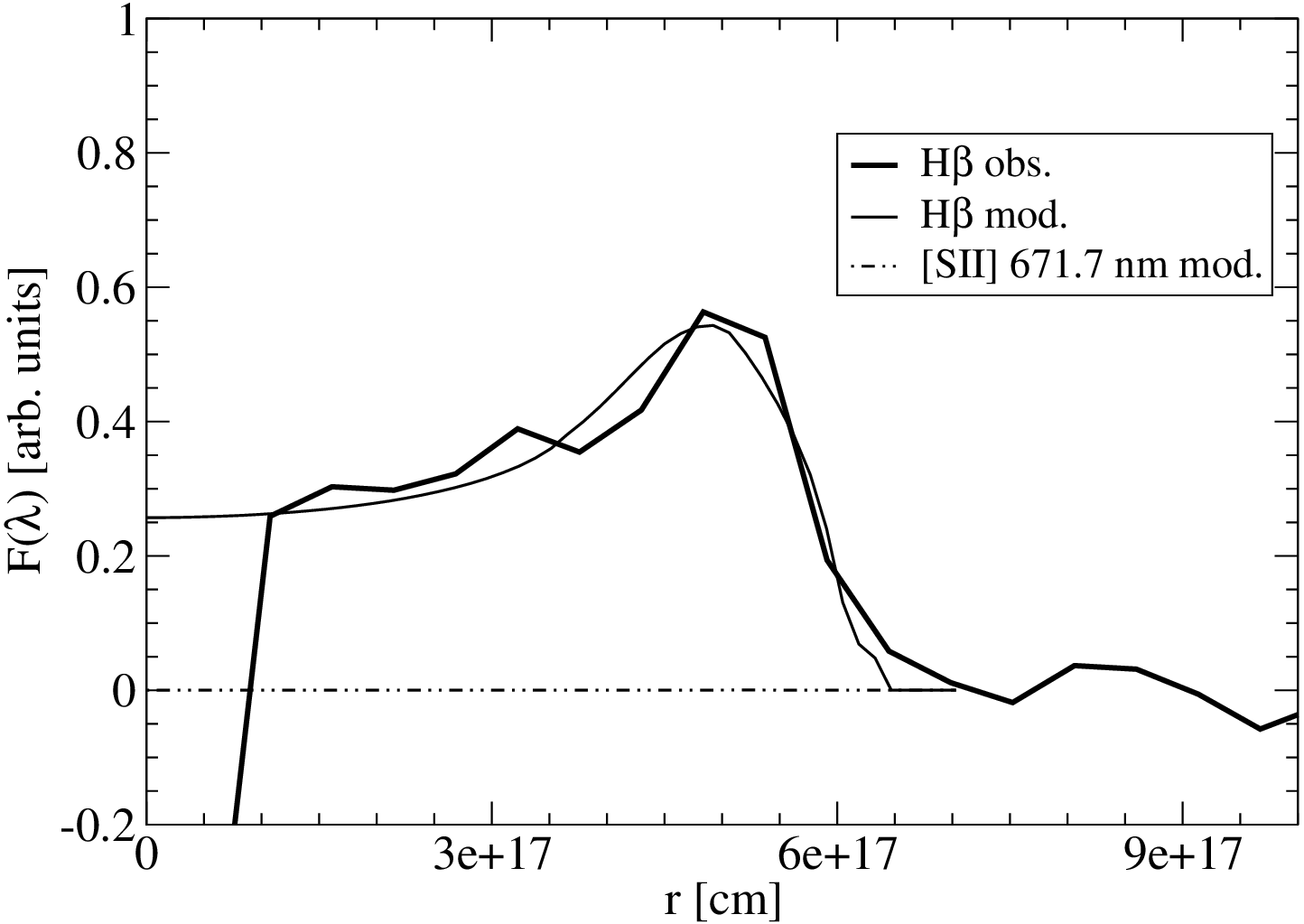}}}

\caption{Comparison of the observation and the model of several lines of MeWe 1-3 (description see Fig. ~\ref{moda15}).}  \label{mod13}
\end{figure*}

\begin{figure*}[ht]
\centerline{\resizebox{18cm}{!}{\includegraphics{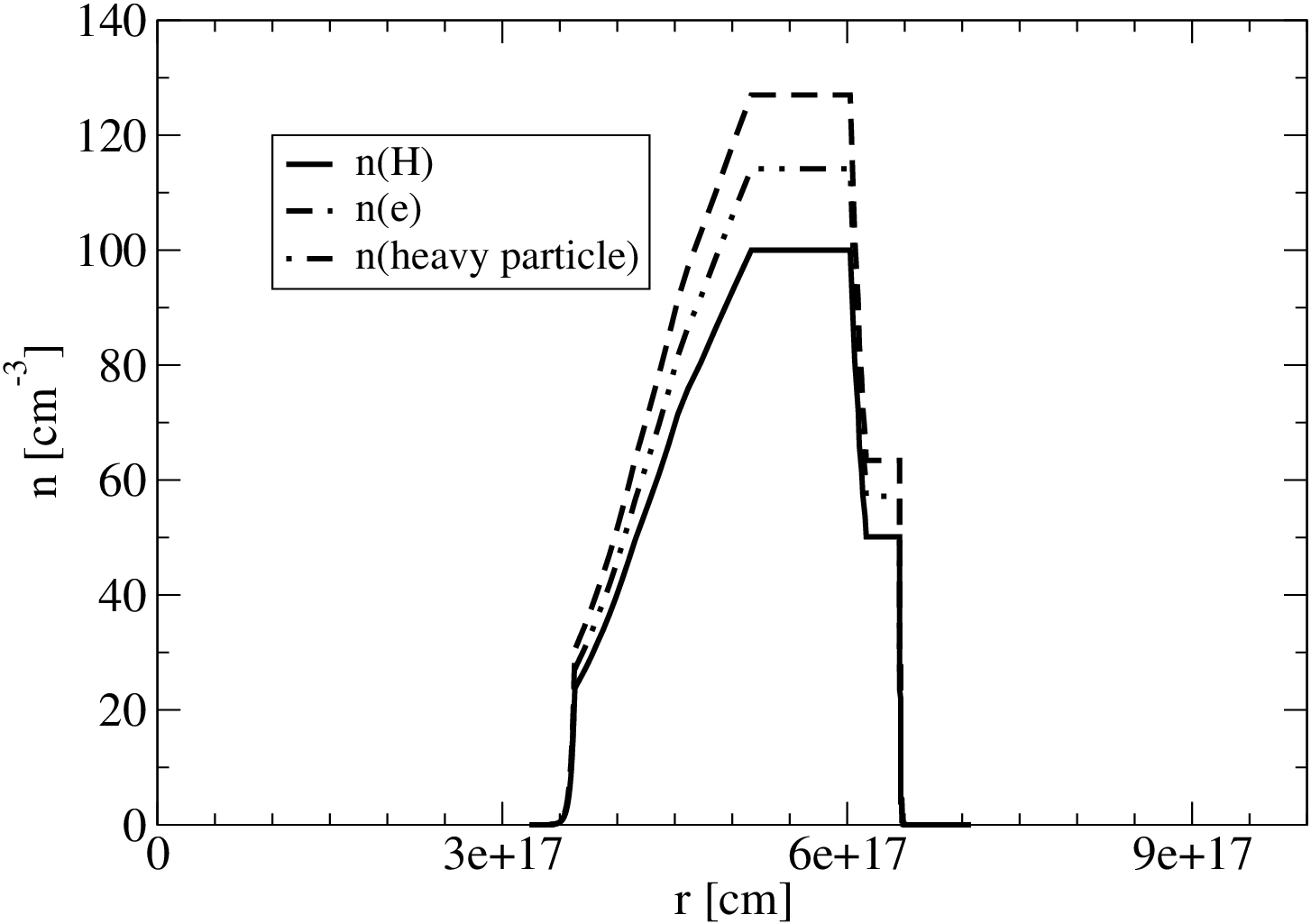}\phantom{lkfag}\includegraphics{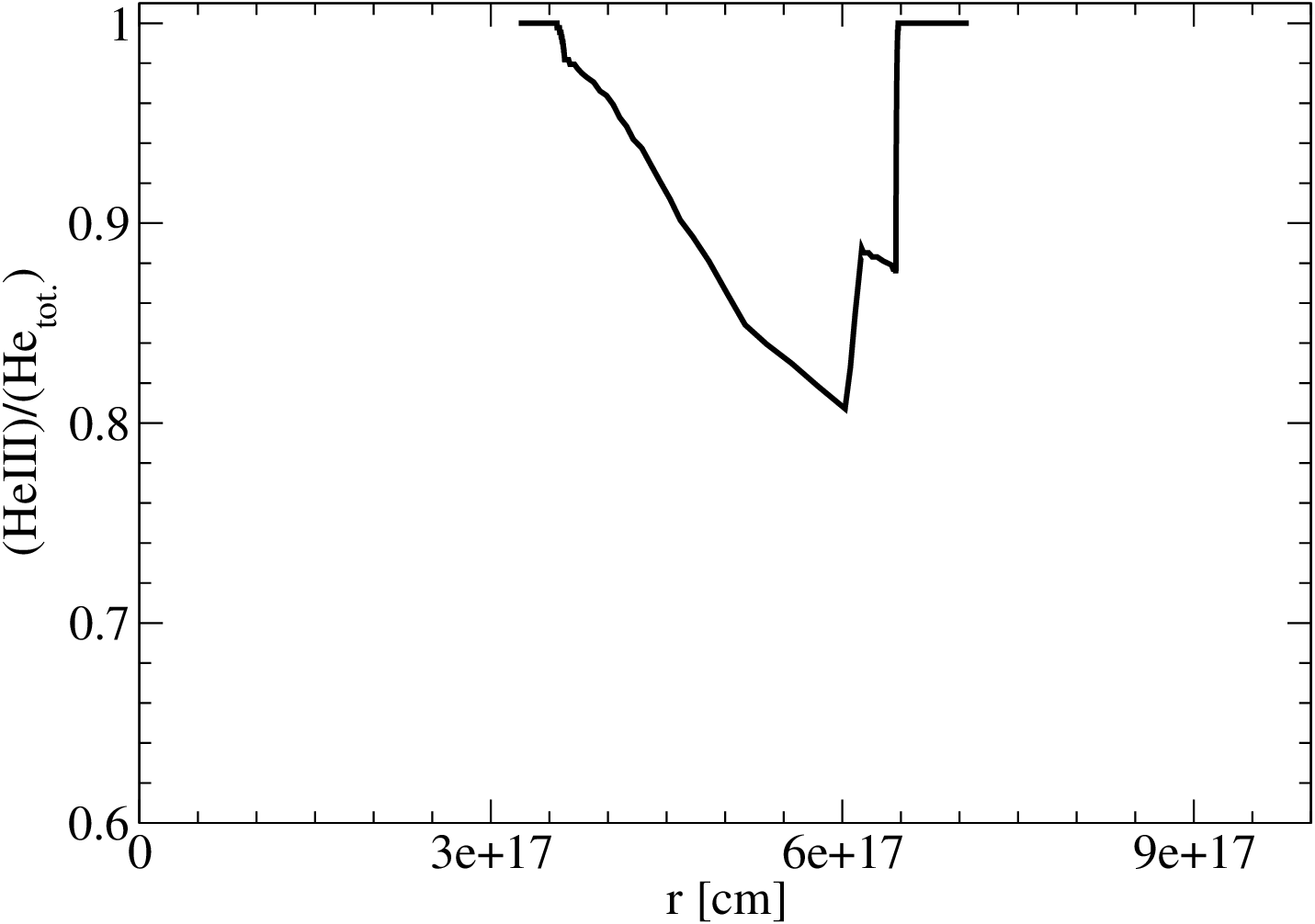}}}

\caption{Density profile (left) and fraction of doubly ionized
helium of MeWe 1-3.}  \label{den13}
\end{figure*}

MeWe 1-3 shows, even if it is not so pronounced like the one in A
20, an inner central cavity. Also in respect to other conditions
it looks like A 20, as there are the similarity of the HeII and
the $\rm H\beta$ line, the relative line strength of [OIII] and
the lack of any [SII] and [NII] lines. Only there is hardly any
halo
visible in MeWe 1-3.\\
Since the emissivity distribution of MeWe 1-3 looks like the one
of A 20 their density profile is similar -- an empty inner part,
and a rapid increase of the density to a plateau, here at 100
H~cm$^{-3}$. Just the narrowness of the halo of only 0.013 pc is
different. In this case the whole nebular hydrogen and helium is
mostly bare of all electrons too.  \\
Also the abundances of the elements are very similar to the one of
A~20. But the average ionization stage is closer to the one of A
15. Also the electron temperature ($15\,800$~K observation,
$22\,500$~K model) is in between the two other nebulae, and that
despite the lower $\rm T_{eff}$ of its central star. The larger
luminosity of the CSPN of MeWe~1-3 and the smaller radius of the
nebula itself is able explain this effect.

\begin{table}[ht]
\caption{Chemical composition of PN MeWe 1-3 (same as in
Tab.~\ref{chem214}).} \label{chem13}
\begin{tabular}{lccc}
\hline\noalign{\smallskip}
& abundance & abundance compared & dominating \\
& & to solar abundance & ionization level \\
\hline \hline\noalign{\smallskip}
He & -0.85 & 0.15 & He III \\
C & -4.45 & -1.0 & C IV \\
N & -5.03 & -1.0 & N IV \\
O & -4.13 & -1.0 & O IV \\
Ne & -5.03 & -1.11 & Ne IV \\
Mg & -5.42 & -1.0 & Mg III \\
Si & -5.45 & -1.0 & Si V \\
S & -5.67 & -1.0 & S V \\
Ar & -6.48 & -1.0 & Ar IV \\
\hline
\end{tabular}
\end{table}

Since the errors of the properties of the objects, given in the
literature, are relatively large, especially for the luminosity
and the distance, we varied them within the ranges of error. The
obtained data are given in Tab.~\ref{datan}.

\begin{table}[ht]
\caption{Obtained basic data of the PNe modeled.} \label{datan}
\begin{tabular}{lcccc}
\hline\noalign{\smallskip}
Name & $T_{eff}$ [kK] & L [$L_{\odot}$] & radius [pc] & dist. [kpc] \\
\hline \hline\noalign{\smallskip}
A 20 & 140 & 3000 & 0.40 & 2.35\\
A 15 & 120 & 4200 & 0.32 & 4.01\\
MeWe 1-3 & 110 & 4500 & 0.21 & 3.95\\
\hline
\end{tabular}
\end{table}

It seems that the temperature and thus the luminosity of the CSPN
of A~20 are underestimated. The uncertain distances of A~20 and
MeWe~1-3 are always smaller in the models compared to the one
given in the literature, but still within the error-deviation.

\section{Discussion and Conclusion}

\begin{figure}[ht]
\centerline{\resizebox{8.8cm}{!}{\includegraphics{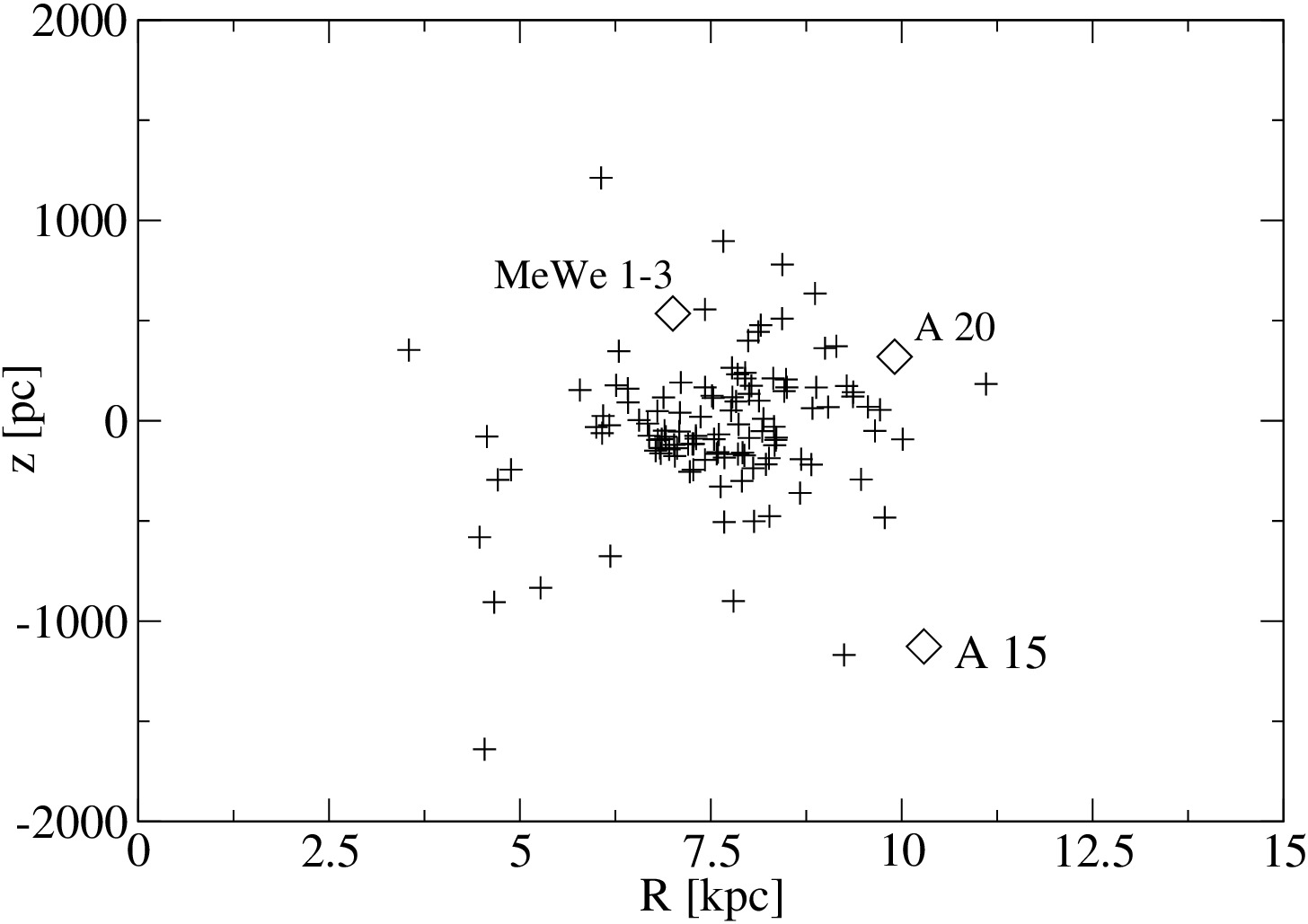}}}

\caption{Here the galactic height $z$ versus the radius from the
galactic center is displayed. The radius of the sun was assumed to
be 8 kpc. The + shows the PNe catalogued by Acker et
al.~(\cite{ESO}) and the $\diamondsuit$ the PNe A 20, A 15 and MeWe
1-3.}  \label{dist}
\end{figure}

The metallicities of all three PNe investigated are lower by a
factor of 10 to 30 compared to the solar abundances, only helium
is little bit more present than in the Sun. Soker~(\cite{soker})
already reports an under-abundance of metals in round PNe.\\
Looking at the galactic coordinates and the distances of the
planetary nebulae one can easily verify that the height above the
galactic plane ($z$) of these nebulae is comparatively large (Fig.~\ref{dist}).\\
The nebulae investigated are located at a $z$ of $320$~pc,
$-1127$~pc and $535$~pc for A~20, A~15 and  MeWe~1-3 respectively.
To compare these values with other one the PNe catalogued by Acker
et al.~(\cite{ESO}) were taken as a representative sample for all
PNe. Half of these planetary nebulae are located below a $z$ of
162~pc. Their average $z$ amounts to 288~pc. The large $z$ of the
three objects investigated indicates that they are so called thick
disk objects and therefore their progenitor was an underabundant
object as well.

The high electron temperature of the nebulae is also linked to the
low metallicity (Stasinska~(\cite{stasi})). As Frank \& Mellema~(\cite{frank}) showed one
important cooling mechanisms in ionized gas is the radiation
emitted by collisionally excited lines. The most important role
plays the [NII] lines in low excited nebulae and the [OIII] lines
in high excited one. Since no [NII]- and less intensive [OIII]
lines, compared to common PNe, are detected a higher temperature
is feasible.

As mentioned above a filling  factor ($\varepsilon$) of 0.1 was
adopted. Since radio data are available for A~20 and A~15 (Condon
\& Kaplan~\cite{con}) it was possible to check this assumption.
Adopting a optically thin wind in the radio regime and therefore
$\rm I_{\nu}\propto \nu^{-0.1}$ allows us to compute the radio
flux at 5 GHz from the 1.4~GHz data and thus the total ionized
mass according Cahn et al.~(\cite{Dist2}).
$$
\rm M_{ion.}=\sqrt{2.266\times 10^{-21}\cdot D^5\cdot\theta^3\cdot F_{5\,GHz}}
$$
where $D$ is the distance in pc, $\theta$ the angular radius in
arcseconds and $\rm F_{5\,GHz}$ the radio flux at 5 GHz in Jansky.
We compared this mass with the one of our model, computed by
$$
\rm
M=4\pi\mu\int_{0}^{+\infty}n_{hp}(r)r^{2} dr
$$
$\mu$ is the mean atomic weight and $\rm n_{hp}(r)$ the heavy
particle density as function of the radius. This results in
$\varepsilon$. Although the error of the obtained filling factor
is quiet high it is clearly smaller than 0.3. Using the sample and
the formula of Mallik \& Peimbert~(\cite{fill2}) we obtain a value
even below 0.05. The formula of Boffi \&
Stanghellini~(\cite{fill3}) give a value of 0.096. But especially
the sample shows no PN with a radius above 0.2~pc and an filling
factor above 0.1. Tests with CLOUDY show that the nebulae are
always optically thin for the UV radiation up to $\varepsilon =
0.5$.\\ Further calculations showed that even stars with
luminosities down to 160~$L_\odot$ are able to ionize most of the
hydrogen and thus a recombination of the shell seems to be
impossible. This is in clear contradiction to the predictions of
the 1D hydro simulations (e.g. Sch\"onberner \& Steffen
\cite{schon}).

While the luminosities of the CSPN found from the investigation of
the PN shells here are about the same as those found {with the
NLTE models of the CSPN spectroscopy}. Our temperatures are -
still within the error bars - slightly higher than those from the
CSPN spectra. As shown by Lechner \& Kimeswenger (\cite{lech}) and
Emprechtinger et al. (\cite{empr}) these parameters are very
sensitive to the line ratios of the highly ionized species.  We
thus suggest this effect to be real. The distances $D_{CSPN}$ have
to be significantly lower than those found by other investigations
in the past.

\begin{acknowledgements}
This research was supported by the DLR under grant 50\,OR\,0201
(TR). We thank the referee R. Szczerba for the valuable comments
improving the original manuscript.
\end{acknowledgements}


\begin{thebibliography}{}

\bibitem[1955]{Abell}
Abell, G.O., 1955, PASP, 67, 258

\bibitem[1992]{ESO}
Acker, A., Ochsenbein, F., Stenholm B. et al., 1992, Strasbourg-ESO Catalogue of Planetary Nebulae

\bibitem[2002]{birgit}
Armsdorfer, B.,  Kimeswenger, S., \& Rauch, T. 2002, RevMexAA
(Serie de Conf.), 12, 180

\bibitem[1995]{blocker}
Bl\"ocker, T., 1995, A\&A, 299, 755

\bibitem[1994]{fill3}
Boffi, F.R., \& Stanghellini, L., 1994, A\&A, 284, 248

\bibitem[1992]{Dist2}
Cahn, J.H., Kaler, J.B., \& Stanghellini, L., 1992, A\&AS, 94, 399

\bibitem[1998]{con}
Condon, J.J., \& Kaplan, D.L., 1998, ApJS, 117, 361

\bibitem[1998]{dwar}
Dwarkadas, V., \& Balick, B., 1998, ApJ, 497, 267

\bibitem[2004]{empr}
Emprechtinger, M., Forveille, T., \& Kimeswenger, S. 2004, A\&A, 423, 1017

\bibitem[1996]{cloudy}
Ferland, G., 1996, A Brief Introduction to Cloudy 90.05, Univ.
Kentucky, Department of Physics and Astronomy, Internal Report

\bibitem[1994a]{frank2}
Frank, A., \& Mellema, G., 1994a, ApJ, 430, 800

\bibitem[1994b]{frank}
Frank, A., \& Mellema, G., 1994b, A\&A, 289, 937

\bibitem[1996]{abund}
Grevesse, N., Noels, A., \& Sauval, A.J. 1996, ASP Conf. Ser., 99,
117

\bibitem[1992]{hamuy}
Hamuy, M., Walker, A.R., Suntzeff, N.B. et al., 1992, PASP 106,566

\bibitem[1984]{heber}
Heber, U., Hunger, K., Jonas, G., \& Kudritzki, R. P., 1984, A\&A,
130, 119

\bibitem[2001]{KS}
Kimeswenger, S. 2001, Rev. Mex. A\&A, 37, 115

\bibitem[1978]{kwok}
Kwok, S., Purton, C.R., \& Fitzgerald, P.M., 1978, ApJ, 219, 125

\bibitem[2004]{lech}
Lechner, M.F.M., \& Kimeswenger, S. 2004, A\&A in press
(astro-ph/0406507)

\bibitem[1988]{fill2}
Mallik, D.C.V., \& Peimbert, M., 1988, RMxAA, 16, 111

\bibitem[1997]{McC}
McCarthy, J.K., Mendez, R.H., \& Kudritzki, R.P., 1997, IAUS, 180,
120

\bibitem[1990]{MeWe}
Melmer, D., \& Weinberger, R., 1990, MNRAS, 243, 236

\bibitem[1988]{mendez}
Mendez, R.H., Kudritzki, R.P., Herrero, A., Husfeld, D., \& Groth,
H.G., 1988,  A\&A, 190, 113

\bibitem[1997]{rauch2}
Rauch, T., 1997, A\&A, 320, 237

\bibitem[2003]{rauch3}
Rauch, T., 2003, A\&A, 403, 709

\bibitem[2004]{rauch4}
Rauch, T., 2004, http://astro.uni-tuebingen.de/$\sim$rauch/

\bibitem[1999]{rauch1}
Rauch, T., K\"oppen, J., Napiwotzki, R., \& Werner, K., 1999,
A\&A, 347, 169

\bibitem[1997]{saurer}
Saurer, W., Werner, K.,  \& Weinberger, R., 1997, A\&A, 328, 598

\bibitem[1979]{Sava}
Savage, B.D., \& Mathis, J.S., 1979, ARA\&A 17, 73

\bibitem[2002]{schon}
Sch\"onberner, D., \& Steffen, M., 2002, RMxAC, 12, 144

\bibitem[1997]{soker1}
Soker, N., 1997, ApJS, 112, 487

\bibitem[2002]{soker}
Soker, N., 2002, A\&A, 386, 885

\bibitem[1978]{stasi}
Stasinska, G., 1978, A\&AS, 32, 429

\bibitem[1998]{steffen}
Steffen, M., Szczerba, R., \& Sch\"onberner, D., 1998, A\&A, 337,
149

\bibitem[1992]{tyl92}
Tylenda, R., Acker, A., Stenholm, B., \& Koeppen, J., 1992, A\&AS,
95, 337

\bibitem[2003]{Tyl}
Tylenda, R., Sidmiak, N., Gorny, S. K., Corradi, R. L. M., \&
Schwarz, H. E., 2003, A\&A, 405, 627


\end{thebibliography}
\end{document}